\documentclass[twocolumn,10pt,prl,showpacs]{revtex4-1}

\usepackage{hyperref}
\usepackage{epsfig}
\usepackage{amsmath,amssymb}
\usepackage{verbatim}

\newcommand{\beq}{\begin{equation}}
\newcommand{\eeq}{\end{equation}}
\def\bea#1\eea{\begin{align}#1\end{align}}
\newcommand{\nn}{\nonumber}

\renewcommand{\d}{\textrm{d}}
\newcommand{\w}{\wedge}
\newcommand{\e}{\textrm{e}}

\def\del {\partial}

\begin{document}

\begin{flushright}
CERN-TH-2018-148
\end{flushright}

\title {\Large\bf On the de Sitter swampland criterion}
 \author{\bf David Andriot}
\affiliation{CERN, Theoretical Physics Department,\\1211 Geneva 23, Switzerland}
\affiliation{\upshape\ttfamily david.andriot@cern.ch}

\begin{abstract}
A new swampland criterion has recently been proposed. As a consequence, it forbids the existence of de Sitter solutions in a low energy effective theory of a quantum gravity. However, there exist classical de Sitter solutions of ten-dimensional (10d) type II supergravities, even though they are unstable. This appears at first sight in contradiction with the criterion. Beyond possible doubts on the validity of these solutions, we propose two answers to this apparent puzzle. A first possibility is that the known 10d solutions always exhibit an energy scale of order or higher than a Kaluza--Klein scale, as we argue. A corresponding 4d low energy effective theory would then differ from the usual consistent truncations, and as we explain, would not admit a de Sitter solution. This would reconcile the existence of these 10d de Sitter solutions with the 4d criterion. A second, alternative possibility is to have a refined swampland criterion, that we propose. It forbids to have both the existence and the stability of a de Sitter solution, while unstable solutions are still allowed.
\end{abstract}

\maketitle

\section{I. Introduction}

Despite remarkable improvements in recent cosmological measurements in terms of precision and data, a plethora of different models remain compatible with observations. Having theoretical criteria assigning a quantum gravity origin to few models while discarding the others would provide a new helpful manner to distinguish between these models. In this paper, we focus on a new proposal for such a criterion.  We provide important clarifications on its interpretation, give tools and ideas to test it, and present a refined version of this criterion.

\subsection{A. Context}

Consider a four-dimensional (4d) gravitational theory, minimally coupled to scalar fields $\phi_i$, governed by a potential $V(\phi_i)$
\beq
{\cal S} = \int \d^4 x \sqrt{|g_4|} \left({\cal R}_4 + \mbox{kin. terms} - V \right) \ , \label{action}
\eeq
where $\mbox{kin. terms}$ denote the scalars kinetic terms, and we choose for convenience units where the 4d Planck mass is set to $1$. To simplify the discussion, we do not include extra content in \eqref{action}, even though the results could easily be adapted. Solutions of this theory \eqref{action} with constant scalar fields would be determined by the following equations of motion
\beq
{\cal R}_{\mu\nu} - \frac{g_{\mu\nu}}{2} {\cal R}_4 = -\frac{g_{\mu\nu}}{2} V  \Rightarrow  {\cal R}_4 = 2 V \ , \quad  \del_{\phi_i} V = 0 \ .\label{eom}
\eeq
Such extrema of the potential, where we denote the value $V|_0$, can then provide solutions with a maximally symmetric 4d space-time, with cosmological constant $\Lambda = \tfrac{1}{2}V|_0$. In particular, de Sitter solutions would have $V|_0 >0$.

Recently, it has been proposed \cite{Obied:2018sgi} that any theory of the form \eqref{action}, not lying in the swampland, should verify the criterion
\beq
|\nabla V| \geq c \, V \ , \label{crit}
\eeq
where $c>0$ and we understand $|\nabla V|$ as $|\nabla V|= \sqrt{g^{ij} \del_{\phi_i} V \del_{\phi_j} V}$, with the field space metric $g_{ij}(\phi)$, readable from the kinetic terms. In other words, any theory \eqref{action} that is a low energy effective theory for a consistent quantum gravity theory should verify the criterion \eqref{crit}. This has the crucial implication that the 4d theory would not admit a de Sitter solution, meaning an extremum of the potential where the 4d space-time is de Sitter. Indeed, the criterion implies at an extremum that $V|_0 \leq 0$, i.e.~it only allows for a Minkowski or anti-de Sitter space-time among maximally symmetric ones.

This ``de Sitter swampland criterion'' \eqref{crit} is motivated by many examples in string theory constructions where similar conditions to \eqref{crit} have been found. It is well-known that many (supersymmetric) Minkowski or anti-de Sitter solutions have been found in string theory, but de Sitter ones are difficult to obtain (we refer to \cite{Bena:2017uuz, Danielsson:2018ztv} for recent reviews on this topic). For de Sitter, two (possibly related) approaches have been followed. The first one consists in looking directly for 10d solutions where the space-time is a product of 4d de Sitter and 6 compact space dimensions. This approach is usually pursued in the framework of ten-dimensional (type II) supergravities, viewed as low energy approximations of string theory, and the solutions then correspond to classical (perturbative) string backgrounds. The compactness of the extra dimensions allow in principle to connect, through dimensional reduction, to a 4d effective theory of the form \eqref{action}, where de Sitter extrema should match the 10d solutions. The second approach is purely 4d: one considers a 4d theory of the form \eqref{action}, argued to come from string theory, and one looks directly for de Sitter solutions at this 4d level. The difficulty of finding solutions in either approach has often been described by conditions analogous to \eqref{crit}, hence the motivation for such a swampland criterion \cite{Obied:2018sgi}.

\subsection{B. Puzzle}

What has been notoriously difficult in string theory is actually to obtain de Sitter {\it vacua}. There is an important distinction to be made with an extremum or solution: a vacuum is a (local) minimum of $V$, i.e.~a (meta)stable solution. Requiring stability adds a further constraint which appears often incompatible with the existence of solutions, already difficult by itself.

Concretely, stability means that the diagonalised mass matrix has only strictly positive entries; there is in particular no tachyon. Provided $g_{ij}$ is positive definite, one should study the sign of the eigenvalues of $\del_{\phi_i} \del_{\phi_j} V|_0$. Sylvester's criterion (see e.g.~\cite{Shiu:2011zt}), applied multiple times when reshuffling lines and columns of the matrix $\del_{\phi_i} \del_{\phi_j} V|_0$, provides a necessary condition for stability, that is $\del^2_{\phi_i} V|_0 > 0$; we will use this in Section IV.

The reason why de Sitter vacua of string theory are so much debated is two-fold: first, there is up-to-date no known example of a 10d classical de Sitter solution that is also metastable. Such solutions have been shown to be very constrained, but at the same time, it has not been possible to fully exclude them. Secondly, de Sitter vacua obtained in the purely 4d approach are under debate, because of the difficulty in lifting them to a controlled string theory construction \cite{Danielsson:2018ztv}.

Given this situation, the criterion \eqref{crit} may look surprising because it forbids de Sitter solutions or extrema, instead of forbidding (only) de Sitter vacua. This may even be problematic because there exist 10d classical de Sitter solutions, which are however tachyonic (we describe them in Section II). If we were arguing on 4d tachyonic de Sitter solutions, one could just question the validity of the 4d theory and classify it as being in the swampland: this is the point of the criterion. On the contrary, classical 10d solutions are, comparatively, much easier to connect to string theory, their embedding in a consistent quantum gravity is harder to question.

The existence of these de Sitter solutions thus appears to be in contradiction with the de Sitter swampland criterion \eqref{crit}. In the remainder of this paper, we propose three different answers to this puzzle, none of them being however definite.

\section{II. Answer 1: the 10d de Sitter solutions are not trustable}

Classical 10d de Sitter solutions have been found in type II supergravities, allowing for all fluxes and including $D_p$-branes and orientifold $O_p$-planes. The 6d compact manifolds on which those solutions have been found are group manifolds, meaning manifolds built out of a 6d Lie group, sometimes divided by a discrete subgroup providing compactness (the lattice). Almost all solutions \cite{Caviezel:2008tf, Flauger:2008ad, Danielsson:2009ff, Danielsson:2010bc, Danielsson:2011au} have been found in type IIA with intersecting $O_6$ (and possibly $D_6$), also viewed as orbifold actions: these solutions are summarized in \cite{Danielsson:2011au}. The only exception is a solution in type IIB \cite{Caviezel:2009tu} with intersecting $O_5$ and $O_7$. Overall, these solutions are non-trivial as they are obtained on non-Ricci flat compact manifolds, and include many fluxes as well as intersecting $D_p/O_p$ sources. It is however at this complexity cost that classical de Sitter solutions can be found, and pass all constraints and no-go theorems \cite{Andriot:2017jhf}.

A first way to enforce the de Sitter swampland criterion \eqref{crit} is to doubt on the validity of these solutions. Several arguments against them have been pointed out \cite{Danielsson:2018ztv}. Let us discuss here only one, often put forward, that is the presence of intersecting $O_p$. The fact they intersect prevents one from describing their backreaction properly, contrary to parallel or even single sources. For the latter, one typically has a warp factor in the metric that accounts for the backreaction; the Laplacian of this warp factor gives rise to a $\delta$-function localizing the source in its transverse directions. In the intersecting case, it is hard to include any such function in the metric that would verify the equations of motion and Bianchi identities (see e.g.~\cite{Smith:2002wn, Arapoglu:2003ah}). Functions are then traded for constants, and as a consequence, the $\delta$-function is replaced by its integrated value over the transverse directions. This is often viewed as ``smearing'' the sources over those directions. Smearing an orientifold is certainly problematic as it should remain at a fixed locus.

This smearing interpretation can however be discussed. The sources are still considered to be along some specific directions, since for instance, the orientifold projection is used. Technically, what is done is integrating equations along some (or all) the 6d directions (see e.g.~\cite{Andriot:2016xvq} for trading the integration over transverse directions to that on 6d, better defined). This integration erases any dependence on internal coordinates and replaces the $\delta$-function by a constant. One then looks for solutions with constant coefficients, actually very suited to group manifolds. Instead of the smearing interpretation, one may rather view this as studying the integrated equations. In addition, those would typically match the 4d equations obtained from the potential, precisely because the potential is also derived by integrating over the 6d space.

If one finds such a solution with constant coefficients, it is of course not guaranteed to have a ``localized'' version \cite{Blaback:2010sj, Junghans:2013xza}. For single or parallel sources, it remains remarkable that the warp factor dependence, together with that of the dilaton, can be completely removed from the equations, leaving only terms without derivatives of those functions \cite{Andriot:2016xvq}, as if one had integrated. But the same does not have to hold for the intersecting case. What is rather believed is that the localized supergravity description is impossible in that case, but the problem would be cured by string theory. Intersecting branes or orientifolds certainly appear in different contexts, such as particle physics models or holographic supergravity backgrounds, where a stringy treatment is sometimes provided. The fact that intersecting $O_6$ can be viewed as orbifolds is also in favour of a more stringy description. It could still happen that a properly treated backreaction would eventually destroy completely the supergravity-approximated solution, but this remains to be checked. In the following, we rather trust these solutions and move on to other possibilities regarding our initial puzzle.

\section{III. Answer 2: the corresponding 4d low energy effective theories do not admit de Sitter solutions}

We present here a new point that could interestingly reconcile the de Sitter swampland criterion \eqref{crit} with the existence of the 10d classical de Sitter solutions. To actually compare those two, one should first fill a gap: what is the 4d theory corresponding to the 10d solutions?

To study the stability of these solutions (and conclude on tachyons), a 4d theory has been needed in the first place, and the one used is a 4d ${\cal N}=1$ gauged supergravity (see e.g.~\cite{Danielsson:2011au}). As first proposed by Scherk and Schwarz \cite{Scherk:1979zr}, group manifolds provide an interesting truncation of the 10d fields to a finite set of modes, given by the left-invariant forms, i.e.~essentially Maurer-Cartan forms with constant coefficients. Since the 10d classical de Sitter solutions are obtained on group manifolds with fields having constant coefficients, this truncation is perfectly suited. The 4d theory resulting from this Scherk--Schwarz truncation is known and given by a 4d gauged supergravity, where the gaugings are in particular the structure constants of the underlying Lie algebra. The Scherk--Schwarz truncation is thought to be a consistent truncation: this means that a solution to the 4d theory can be lifted to a solution of the 10d theory. One way to view this is to say that the finite set of modes is ``independent'' or decoupled from the rest of the 10d modes, and so one can study its physics independently. In addition, because the 10d solutions are only expressed in terms of the same finite set of modes, they will appear as solutions to the 4d theory. In short, thanks to the consistent truncation, the resulting 4d ${\cal N}=1$ gauged supergravity will admit the same de Sitter solution when solving the extrema conditions \eqref{eom}. This is certainly the point in conflict with the criterion \eqref{crit}.

There is however another important ingredient: the notion of swampland, or in other words, that of having a {\it low energy} effective theory. The 4d ${\cal N}=1$ gauged supergravities are derived by a well-defined procedure (a consistent truncation) from a quantum gravity theory, but as we will argue, {\it they are very unlikely to be low energy effective theories}. Indeed, what usually happens in consistent truncations is that one truncates some light modes, or keeps heavier modes than those truncated. We now argue that this should always be the case for the 10d classical de Sitter solutions: they always exhibit an energy scale of order or higher than a Kaluza--Klein scale, while you need to truncate the latter at low energy. We describe the theory one would rather obtain after a low energy truncation.\\

In recent works \cite{Andriot:2016rdd, Andriot:2018tmb},  we determined and studied the spectrum of the Laplacian on a specific group manifold, the 3d Heisenberg nilmanifold. This manifold is built out of the Heisenberg algebra and therefore has only one structure constant, $f^3{}_{12}$. This constant should be quantized for geometric reasons \cite{Andriot:2016rdd}, related to compactness and the lattice action. In the basis where the internal metric is $\delta_{ab}$, the radii $r^{m=1,2,3} > 0$ of the three directions enter the structure constant, then given by
\beq
f^3{}_{12} = \frac{r^3 N}{r^1 r^2} \ ,\ N \in \mathbb{Z}^* \ .
\eeq
The structure constant is also generically related to the Levi-Civita spin connection through e.g.~$f^a{}_{bc} = 2 \omega_{[b}{}^a{}_{c]}$. The Ricci scalar can then be expressed purely in terms of $f^a{}_{bc}$. We then understand that there are typically two energy scales on group manifolds: those given by the radii, $1/r^m$, and the scales given by the curvature or structure constants. In \cite{Andriot:2018tmb}, we proposed a small fiber/large base approximation given by
\beq
r^3 \leq |N| r^3 \ll r^1, r^2\ \Rightarrow \ |f^3{}_{12}| \ll \frac{1}{r^1} , \frac{1}{r^2} \ll \frac{1}{r^3} \ . \label{lowenergyapprox}
\eeq
This approximation generates an interesting hierarchy between the curvature scale and the radii scale, or in supergravity terms, between the geometric flux and the Kaluza--Klein scale. Having the explicit Laplacian spectrum, we could show that truncating, thanks to this approximation, the modes heavier than $|f^3{}_{12}|$, one would be left with only a finite set of light modes: this is then a {\it low energy} truncation. Remarkably, this finite set of light modes turned out to correspond to the left-invariant forms, i.e.~the set of modes one would keep in the Scherk--Schwarz truncation. On a nilmanifold, there is therefore a chance, rare otherwise, that the consistent truncation is also a low energy truncation, and the 4d gauged supergravity a low energy effective theory. As discussed in \cite{Andriot:2018tmb}, few more steps have to be taken before proving this, in particular because the geometric scales are not the only ones entering the game: a concrete background with other fluxes could modify the hierarchy. The fact that the Kaluza--Klein towers of the Laplacian spectrum are truncated in this way is still a good start.

However, {\it no classical de Sitter solution is known on a nilmanifold}. And turning to the other group manifolds will completely change the situation. Nilpotent algebras are such that there exists a basis where for $a\neq b$, if $f^a{}_{bc} \neq 0$, then $f^b{}_{ad} = 0$ $\forall d$ (see e.g.~the list of 6d algebras in \cite{Grana:2006kf}). This property is related to the topology of these manifolds, made of an ordered succession of circle fibrations. On the contrary, any other Lie algebra would have, at least in some basis, some structure constant for which the contrary would hold: they would admit for instance both $f^3{}_{12} \neq 0$ and $f^1{}_{32} \neq 0$. It is clear for semi-simple algebras that can always be written in a basis where structure constants are fully antisymmetric, and one can also verify on the examples of solvable Lie algebras in \cite{Andriot:2010ju}. Because the Maurer-Cartan one-forms are such that $\d e^a = -\frac{1}{2} f^a{}_{bc} \e^b \w e^c$ and $e^a= e^a{}_m \d y^m$ where $ e^a{}_m $ is the vielbein and $y^m$ the internal coordinates, the scaling of $f^a{}_{bc}$ with the radii is then always the same. We obtain
\beq
f^3{}_{12} = \frac{r^3 N}{r^1 r^2} \ ,\ f^1{}_{32} = \frac{r^1 N'}{r^3 r^2} \ ,
\eeq
for any group manifold being not a nilmanifold. The numbers $N$ and $N'$ need not be integers but are quantized in different manners, depending on the algebra and the lattice (see e.g.~\cite{Andriot:2010ju, Grana:2013ila, Andriot:2015sia} for solvmanifolds). We now see that the low energy approximation \eqref{lowenergyapprox} will not work here: if $r^3\ll r^1$, $|f^3{}_{12}|$ is small but $|f^1{}_{32}|$ becomes large, compared to $1/r^2$. This is problematic, because one typically needs to truncate at the Kaluza--Klein scale, here $1/r^2$, to avoid an infinite tower of modes. Doing so, one would truncate the modes of energy scale given by $|f^1{}_{32}|$, thus effectively erasing that structure constant. Another option is to set all radii to be of the same order, but there is then no internal scale separation anymore: the structure constants are of the same order as the Kaluza--Klein scale, and may as well be truncated, leaving only massless modes. We conclude that any group manifold, different than a nilmanifold, provides energy scales of order or higher than a Kaluza--Klein scale. Equivalently, the Scherk--Schwarz truncation, that keeps all left-invariant forms and thus all structure constants contributions, cannot be a low energy truncation away from a nilmanifold (explicit reductions of \cite{Caviezel:2008ik} support this claim). As a consequence, the 4d gauged supergravities, having all structure constants in the potential, cannot be low energy effective theories. Furthermore, {\it the low energy effective theory on a group manifold is necessarily the same as one obtained on a nilmanifold}, i.e.~in the above example, where $f^1{}_{32}$ is truncated and $f^3{}_{12}$ remains alone as in a nilpotent algebra, or where both are truncated.\\

This claim, that remains to be checked by explicit dimensional reductions on concrete backgrounds, reconciles the de Sitter swampland criterion \eqref{crit} with the existence of 10d classical de Sitter solutions on group manifolds. As observed in \cite{Danielsson:2011au} after an important search, {\it there is no known de Sitter solution on a nilmanifold}. In \cite{Andriot:2018ept}, we will provide more analytical evidence of this statement. We now combine the two claims: the string 4d low energy effective theory on any group manifold is that obtained on a nilmanifold, and there is no de Sitter solution on a nilmanifold. We conclude that one cannot have a de Sitter extremum in a 4d low energy effective theory obtained from string theory on a group manifold, despite the existence of 10d solutions on such manifolds: the two are now compatible.

The above reasoning implies that the physics of 10d classical de Sitter solutions cannot be described at a 4d level, because their internal geometry exhibits energy scales which are too high. The value of the cosmological constant in these solutions is however not necessarily big, as it could be fixed by different ingredients, such as the fluxes or even the string coupling constant, potentially providing smaller energy scales. This value should be checked on explicit solutions. Still, if their physics can only be described at a 10d level, it is challenging but also interesting and new to build cosmological models using such solutions.

\section{IV. Answer 3: we require a refined de Sitter swampland criterion}

As recalled in Section I.B., the issue with string theory and de Sitter space-time is not about the existence of solutions, i.e.~an extremum of $V$ (even though this remains difficult) but about having solutions which are (meta)stable, i.e.~a vacuum. One could infer that a criterion inspired by this situation should not be \eqref{crit}, but rather one that involves both difficulties, the existence and the stability, so $V>0$ for de Sitter, $\del_{\phi_i} V = 0$ for the existence and $\del^2_{\phi_i} V > 0$ for the stability (a necessary condition, as argued in Section I. B.). A natural combination is then given by the following criterion, to be satisfied by any low energy effective theory of a quantum gravity theory
\bea
& \exists\ b_i \in \mathbb{R}, c_i \in \mathbb{R}_+ \ \mbox{such that} \label{crit2}\\
& \ V + \sum_i b_i\ \phi_i \del_{\phi_i} V + \sum_i c_i\ \phi_i^2 \del^2_{\phi_i} V \leq 0 \ , \nn
\eea
where by $c_i \in \mathbb{R}_+$, we mean $c_i \geq 0$. On a solution, the condition boils down to
\beq
\mbox{Solution:} \ V|_0 + \sum_i c_i \ (\phi_i^2 \del^2_{\phi_i} V)|_0 \leq 0 \ ,
\eeq
which forbids the possibility of having both a stable solution and a de Sitter solution; it however leaves room for tachyonic de Sitter solutions. Regarding Minkowski or anti-de Sitter solutions, they can be accommodated by this criterion in various cases, for instance if one can choose $c_i=0$, i.e.~prove in general that the theory only allows for $V|_0 \leq 0$: this is for instance the case of supersymmetric solutions in ${\cal N}=1$ supergravity without D-term.

This new ``de Sitter vacua swampland criterion'' \eqref{crit2} should certainly be checked on examples (see e.g.~\cite{Junghans:2016abx} for a realisation) and may find further refinements. For instance, the first derivative term can be replaced by a power of $|\nabla V|$ or any other vanishing combination. For $V>0$, it could then be refined in terms of single field inflation slow-roll parameters towards
\beq
\sqrt{\epsilon_V} - a\ \eta_V \geq c \ ,\qquad {\rm with}\ a \geq 0\ , \ c > 0 \ ,
\eeq
by redefining the scalar field and fixing $b_1 <0$; this is a simple and interesting extension of the original criterion \eqref{crit}. Studying the cosmological implications of the condition \eqref{crit2} would also be interesting, similarly to \cite{Agrawal:2018own}, in particular for multi-field inflation. But the aim for now is simply to indicate that if one wants to propose a swampland criterion related to the current de Sitter situation in string theory, a natural one would look like \eqref{crit2}. At the same time, we presented in Section III an explanation that could reconcile both the de Sitter swampland criterion \eqref{crit} of \cite{Obied:2018sgi} and the existence of 10d classical de Sitter solutions. We hope that this work will motivate further checks of either possibility, or stimulate new proposals.

\section{Acknowledgements}

I wish to thank D.~Roest, G.~Shiu, P.~Tourkine, D.~Tsimpis, T.~Van Riet, T.~Weigand, T.~Wrase and the whole CERN string group for stimulating discussions that motivated me to write this paper.

\end{document}